\documentclass{elsart}
\usepackage[dvips]{graphics}  
\begin{document}
\runauthor{Cicero, Caesar and Vergil}

\begin{frontmatter} 

\title{
A New Parameter $F$ to Classify Cellular Automata Rule Table Space and
a Phase Diagram in $\lambda-F$ Plane
}

\author{Sunao Sakai and Megumi Kanno}

\address{
Faculty of Education, Yamagata University, Yamagata,
990-8560, Japan}
\thanks{E-mail: sakai@e.yamagata-u.ac.jp}

\begin{abstract}
It is shown that for the N-neighbor and K-state cellular automata, the
class II, class 
III and class IV patterns coexist at least in the range 
$\frac{1}{K} \le \lambda \le 1-\frac{1}{K} $.  
The mechanism which determines the difference between the pattern
classes at a fixed $\lambda$ is found, and it is studied
quantitatively by introducing a new
parameter $F$. Using the parameter F and $\lambda$, the phase diagram
of cellular automata is obtained for 5-neighbor and 4-state cellular
automata.\\  

\vspace*{0.3cm}
\end{abstract}
\maketitle
\end{frontmatter}

\section{Introduction}
Cellular automata (CA) has been one of the most 
studied fields in the research of 
complex systems. 
Various patterns has been generated
by choosing the rule tables.
Wolfram\cite{wolfram} has classified these patterns into four rough
categories:
class I (homogeneous), class II (periodic), class III (chaos) and 
class IV (edge of chaos). 
The class IV patterns have been the most interesting target for the
study of CA, because it provides us with an example of the
self-organization in a simple
system and it is argued that the possibility of 
computation is realized by the complexity at the edge of
chaos\cite{wolframs,langton,mitchell}.\\
\indent
A more detailed classification of CA, has been carried out mainly for
the elementary cellular automata (3-neighbor and 2-state
CA)\cite{hanson,wuensche}, in which the pattern is studied quite
accurately for each rule table. 
However the number of rule table in
N-neighbor and K-state cellular automata grows like $K^{K^{N}}$. Therefore
except for a few smallest combinations of the $N$ and $K$, the numbers
of the rule tables become so large that studies of the CA dynamics for
all rule tables are impossible even with the fastest supercomputers. \\
\indent
On the other hand the rule table of edge of 
chaos is rare in the whole CA rule table space, therefore 
it is important to find a set of parameters by which the pattern
classes could be classified, and to determine a
phase diagram of CA in these parameter space, even if it is a
qualitative one.\\
\indent
Langton has introduced $\lambda$ parameter and
argued that as $\lambda$ increases the pattern class
changes from class I to class II and then to class
III. And in many cases, class IV behavior is
observed between class II  and class III 
patterns\cite{langton0,langton,langton2}.
The $\lambda$ parameter represents rough behavior of CA in the rule table
space, but finally
does not sufficiently classify the quantitative behavior of CA.
It is well known that different pattern classes coexist at the same
$\lambda$. 
Which of these pattern classes is chosen, depends on
the random number.
The reason or mechanism for this is not yet known;  
we have no way to control the pattern classes at fixed $\lambda$.
And the transitions between a periodic to
chaotic pattern classes are observed in a rather wide range of $\lambda$.
In Ref.\cite{langton2}, a schematic phase-diagram was sketched. However
a vertical axis was not specified.
Therefore, it is has been thought 
that more parameters are necessary to arrive at a more quantitative
understanding of the rule table space of the CA.\\
\indent
In this article, we will report a mechanism which discriminates the
pattern classes at a fixed $\lambda$. The mechanism is closely related
to the structure of the rule tables and is expressed by the  
numbers of rules which breaks
strings of quiescent state.
For N-neighbor and K-state CA, it is found that in the region
$1/K \le \lambda \le 1-1/K$, the class II, class III and class IV
pattern classes coexist.\\
\indent 
This property is studied quantitatively by
introducing a new parameter $F$, which is taken to be orthogonal to
$\lambda$. 
In the region $1/K \le \lambda \le 1-1/K$, 
the maximum of $F$
correspond to class III rule tables while minimum of $F$, to class II
or class I rule tables. Therefore
the transition of the pattern classes takes place somewhere between
these two limits without fail. 
If we determine the region of $F$, where the transitions of the pattern
classes take place, we could obtain the phase diagram in $\lambda-F$
plane. \\
\indent
The determination of the phase diagram is carried out for
5-neighbor
and 4-state CA. In this case, phase boundary is not sharp but has some
range in $\lambda-F$ plane. The region has a gentle slope
as a function of $\lambda$, and  
extends over the range
$0.2 \le \lambda \le 0.8$. This means
that for this CA the edge of chaos could be found at least in this
range in $\lambda$.\\
\indent
In section 2, we will briefly summarize our notations and present
a key discovery, which leads us to the understanding of the structure
of the rule table and pattern classes. It strongly suggested that the
rules which break
strings of the quiescent states play an important role for
the pattern classes.\\
\indent
In section 3, we classify rule tables according to the
destruction and construction of strings of the quiescent states,
and carry out the replacements of the rules to
change the chaotic pattern class into periodic one 
and vice versa while keeping $\lambda$ fixed.
The reason why the patten classes changes  by the replacements is 
discussed, and we will show that in the region $1/K \le \lambda \le
1-1/K$, the change of the pattern classes takes place without fail
by the replacements. \\
\indent
In section 4, the result obtained in section 3 is studied 
quantitatively by introducing a new parameter $F$.
Using $F$ and $\lambda$, we 
determine the phase diagram in the $\lambda-F$ plane for 5-neighbor
and 4-state CA. \\
\indent
Section 5 is devoted to conclusions and discussions.

\section{Summary of CA and a key discovery}
\subsection{ Summary of Cellular Automata }
In order to make our arguments concrete, we focus mainly on 
the one-dimensional 5-neighbor and 4-state CA in the following,
however, the qualitative conclusions hold true for other CAs.
This point will be discussed in subsections 3.2.\\
\indent
We will briefly summarize our notation of CA\cite{wolfram,langton}.
In our study, the site consists of
150 cells having the periodic boundary condition. The states are denoted
as $s(t,i)$. 
The $t$ represents the time step which takes an integer value,
and the $i$ is the position of cells which range from $0$ to
$149$. 
The $s(t,i)$ takes values $0,1,2,$ and $3$, and the state $0$ is taken
to be the quiescent state.
The set of the states $s(t,i)$
at the same $t$ is called the configuration.  \\
\indent
The configuration at time $t+1$ 
is determined by that of time $t$ by using following local relation,
\begin{equation}
s(t+1,i) = T(s(t,i-2),s(t,i-1),s(t,i),s(t,i+1),s(t,i+2)). 
\label{table}
\end{equation}
The set of the mappings 
\begin{equation}
T(\mu,\nu,\kappa,\rho,\sigma)=\eta,(\mu,\nu,etc.= 0,1,2,3) 
\label{r_table}
\end{equation}
is called the
rule table.  The rule table consists of $4^5$ mappings,
which are selected from a total of $4^{1024}$ elements.\\
\indent
The $\lambda$ parameter is defined as\cite{langton}
\begin{equation}
\lambda=\frac{N_{h}}{1024}, 
\label{lambda}
\end{equation}
where $N_{h}$ is the number in which $\eta$ in Eq.\ref{r_table} is not
equal to $0$. In other words 
the $\lambda$ is the probability that the rules do not select the
quiescent state in next time step. In the following we set the
rule tables randomly with the probability $\lambda$. 
We choose $1024-N_{h}$ rules randomly, and set $\eta=0$ in the 
right hand side of Eq. \ref{lambda}.
For the rest of the $N_{h}$ rules, the $\eta$ picks up the number $1,2,3$
randomly.
The initial configurations are also set randomly.\\
\indent
The time sequence of the configurations 
is called a pattern. The patterns are classified 
roughly into four classes established by Wolfram\cite{wolfram}. 
It has been known that as the
$\lambda$ increases the most frequently generated 
patterns change from homogeneous (class I) to
periodic (class II) and then to chaotic (class III),  
and at the region between class II and class III, the  edge of
chaos (class IV) is located.
\subsection{Correlation between pattern classes and rules 
which break strings of quiescent state at a fixed $\lambda$} 
\begin{figure}
\begin{center}
\scalebox{0.58}{ { \includegraphics{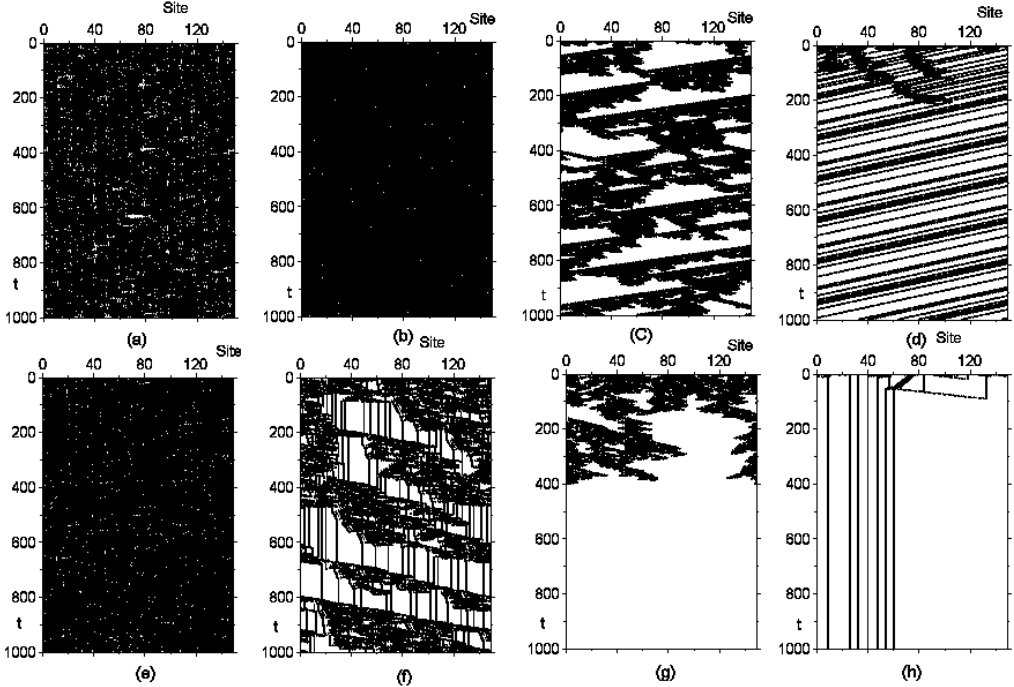} } }
\caption{
The pattern classes at $\lambda=0.44$. The quiescent state is
shown by white dot, while other states are indicated by black
point. Fig.\ref{pattern_at_lambda0.44}(a) corresponds to $N_4=4$,
 Fig.\ref{pattern_at_lambda0.44}(b), Fig.\ref{pattern_at_lambda0.44}(c),
and   Fig.\ref{pattern_at_lambda0.44}(d), to $N_4=3$, 
Fig.\ref{pattern_at_lambda0.44}(e), Fig.\ref{pattern_at_lambda0.44}(f), and
Fig.\ref{pattern_at_lambda0.44}(g), to $N_4=2$, and 
Fig.\ref{pattern_at_lambda0.44}(h) corresponds to $N_4=1$. 
}
\label{pattern_at_lambda0.44}
\end{center}
\end{figure}

In order to find the reason why the different pattern classes are
generated with the same $\lambda$, 
we have started to collect rule tables of 
different pattern classes, and tried to
find the differences between them. 
We have fixed at $\lambda=0.44$ $(N_{h}=450)$,
because at this point the chaotic, edge of chaos, and periodic patterns
are generated with a similar ratio.
By changing the random number, we have
gathered a few tens of the rule tables and classified them
into chaotic, edge of chaos, and periodic ones.\\
\indent 
In this article, a pattern is considered the edge of chaos when its
transient length\cite{langton} is longer
than $3000$ time steps.\\
\indent
First, we study whether or not the pattern classes are sensitive to the
initial configurations.
We fix the rule table and change the initial configurations.
The details of the patterns depend on the initial configurations, but
the pattern classes are not changed\cite{wolfram}.
Thus the difference of the pattern classes is due to the differences 
in the rule tables, and the target of our inquiry has to do with the
differences between them.\\
\indent
For a little while, we do not impose a quiescent
condition (QC)\cite{langton},\\
$T(0,0,0,0,0)=0$,
because without this condition, 
the structure of the rule table becomes more transparent.
This point will be discussed at the footnote 4 in section 4.\\
\indent
After some trial and error, we have found a strong correlation between
the pattern classes and the QC.
For class II patterns, 
the probability of the rule table,
which satisfies the QC is much larger than that of the class III
patterns. This correlation has suggested that the rule 
$T(0,0,0,0,0)=h,\hspace{0.2cm} h \neq 0$, which breaks the string of the 
quiescent states with length 5, 
pushes the pattern toward chaos.
We anticipate that the similar situation will hold for the strings of
quiescent states with length 4.\\
\indent
We go back to the usual definitions of CA. In the following we discuss 
CA under QC, $T(0,0,0,0,0)=0$. 
We study the correlation between the number of the rules of
Eq.\ref{d4} and the pattern classes:
\begin{equation}
T(0,0,0,0,i)=h , \hspace {0.5cm} T(i,0,0,0,0)=h,(i,h=1,2,3).
\label{d4}
\end{equation}
\indent
These rules break length 4 strings of the quiescent states, and
will also push the pattern toward chaos
\footnote{Similar ideas had
been noticed by Wolfram and Suzudo with the
arguments of the unbounded growth\cite{wolfram} and
expandability\cite{suzudo}.\\}.
We denote the total number of rules of Eq.\ref{d4} in a rule table 
as $N_4$.
We have collected 30 rule tables and grouped them by the number $N_4$.
We have 4 rule tables with $N_4 \ge 4$ , 13 rule tables with $N_4=3$, 9
rule tables with $N_4=2$ and 4 rule tables with $N_4 \le 1$.
When $ N_4 \ge 4$, all rule tables generate 
chaotic patterns, while when $N_4 \le 1$, only
periodic ones are generated.
At $N_4=3$ and $N_4=2$, chaotic, edge of chaos, and periodic
patterns coexist.
Examples are shown in the Fig.\ref{pattern_at_lambda0.44}.
The coexistence of three pattern classes at $N_4=3$ is seen in
Fig.\ref{pattern_at_lambda0.44}(b),
Fig.\ref{pattern_at_lambda0.44}(c) and Fig.\ref{pattern_at_lambda0.44}(d)
and that of $N_4=2$ is exhibited in 
Fig.\ref{pattern_at_lambda0.44}(e),
Fig.\ref{pattern_at_lambda0.44}(f) and Fig.\ref{pattern_at_lambda0.44}(g).\\
\indent
The strong correlation between $N_4$
 and the pattern classes has been
observed in this case too, as anticipated.
These discoveries have provided us with a key hint leading us to the
hypothesis 
that the rules, which break strings of the quiescent states, will
play a major role in the pattern classes.

\section{Structure of rule table and pattern classes}
\subsection{Structure of rule table and replacement experiment}
\begin{figure}
\begin{center}
\scalebox{0.58}{ { \includegraphics{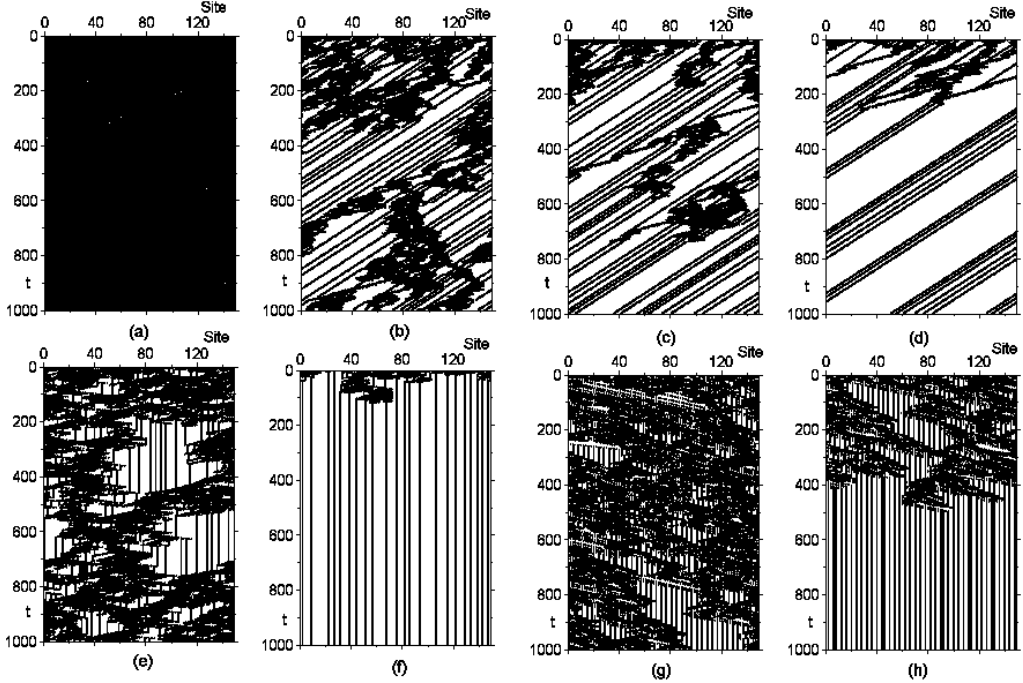} } }
\caption{
Example of the replacement experiments at $\lambda=0.6$.
Fig.\ref{replace1}(a) is obtained randomly with probability
$\lambda=0.6$, by the method explained in subsection 2.1. 
Fig.\ref{replace1}(b) to Fig.\ref{replace1}(h) are obtained by the
replacements of the rules
of Fig.\ref{replace1}(a), which are summarized in Table \ref{table2}.
}
\label{replace1}
\end{center}
\end{figure}

In order to test the hypothesis of the previous section,
we classify the rules into four groups according to the
operation on strings of the quiescent states.
In the following, Greek characters in the rules
represent groups $0,1,2,3$ while Roman, represent groups $1,2,3$.\\
\noindent
Group 1: $T(\mu,\nu,0,\rho,\sigma)=h$.\\
The rules in this group break strings of the quiescent states.\\
Group 2: $T(\mu,\nu,0,\rho,\sigma)=0$. \\
The rules of this group conserve them.\\
Group 3: $T(\mu,\nu,i,\rho,\sigma)=0$.\\
The rules of this group develop them.\\
Group 4: $T(\mu,\nu,i,\rho,\sigma)=l$. \\
The rules in this group do not affect string of quiescent states in next 
time step.\\
\indent
The sum of the numbers of the group 1 and group 2 rules is 256, while that of
group 3 and group 4 rules is 768. The number of each group of rules 
included in the rule table is determined mainly by the probability
$\lambda$, therefore it suffers from fluctuation due to randomness.\\
\indent
The group 1 rules are further classified into five types according to
the length of string of quiescent states, which they break. These are
shown in Table \ref{destruc}. 
\renewcommand{\arraystretch}{0.9}
\begin{table}[h]
\caption{ The classification of the rules in group 1.}
\vspace*{0.3cm}
\label{destruc}
\begin{center}
\begin{tabular}{|c|c|c|c|c|c|c|}
     \hline
     \multicolumn{1}{|c|}{type} &
     \multicolumn{1}{|c|}{Total Number} &
     \multicolumn{1}{|c|}{Name}&
     \multicolumn{1}{|c|}{Replacement}\\
     \hline
        $T(0,0,0,0,0)=h$  &1  &D5  &RP5,RC5\\
     \hline
        $T(0,0,0,0,i)=h$  &3  &D4  &RP4,RC4\\
        $T(i,0,0,0,0)=h$  &3  &    &    \\
     \hline
        $T(0,0,0,i,\sigma)=h$ &12 &   &    \\
        $T(i,0,0,0,m)=h$      &9  &D3 &RP3,RC3\\
        $T(\mu,j,0,0,0)=h$    &12 &   &    \\
     \hline
        $T(\mu,j,0,0,m)=h$    &36 &D2 &RP2,RC2 \\
        $T(i,0,0,l,\sigma)=h$ &36 &   &       \\
     \hline
        $T(\mu,j,0,l,\sigma)=h$ &144 &D1 &RP1,RC1\\
     \hline
\end{tabular}  
\end{center}  
\end{table}
\vspace{0.5cm}
\\
The D5 rule is always excluded from rule tables by the 
quiescent condition.\\
\indent
Our hypothesis presented at the end of the
section 2 is expressed more quantitatively as follows;
the numbers of the D4, D3, D2, and D1 rules shown in Table
\ref{destruc} will mainly determine the pattern classes.\\
\indent
In order to test this hypothesis we artificially change the numbers of
these rules in Table \ref{destruc} while keeping
the $\lambda$ fixed.
For D4 rules, we carry out the replacements defined by the following
equations, 
\begin{equation}
\begin{array}{ll}
T(0,0,0,0,i)=h \rightarrow T(0,0,0,0,i)=0, \\ 
or \hspace {0.1cm}T(i,0,0,0,0)=h \rightarrow T(i,0,0,0,0)=0, \\ 
T(\mu,\nu,j,\rho,\sigma)=0 \rightarrow T(\mu,\nu,j,\rho,\sigma)=l,\\
\label{toperio} 
\end{array}
\end{equation}
where except for $h$, the groups $\mu$, $\nu$, $\rho$, $\sigma$, $j$ and $l$ 
are selected randomly.
Similarly the replacements are generalized for D3, D2, and D1 rules
, which are denoted as RP4 to RP1 in Table \ref{destruc}. 
They change the rules of group 1 to
that of group 2 together with group 3 to group 4 and are expected
to push the rule table toward the periodic direction. \\
\indent
The reverse replacements for D4 are
\begin{equation}
\begin{array}{ll}
T(0,0,0,0,i)=0 \rightarrow T(0,0,0,0,i)=h, \\
or \hspace {0.1cm}T(i,0,0,0,0)=0 \rightarrow T(i,0,0,0,0)=h, \\
T(\mu,\nu,j,\rho,\sigma)=l \rightarrow T(\mu,\nu,j,\rho,\sigma)=0,\\
\label{tochaos} 
\end{array}
\end{equation} 
which will push the rule table toward the chaotic direction. In this case,
the groups $h$, $\mu$, $\nu$, $j$, $\rho$, and $\sigma$ are
selected randomly.
Similarly we introduce the replacements for D3, D2, and D1, which will
be called RC4 to RC1 in the following.\\
\indent
By the replacement of RP4 to RP1 or 
RC4 to RC1, 
we change the numbers of tfhe rules in Table
\ref{destruc} while keeping the $\lambda$ fixed. We denote these
numbers $N_4$, $N_3$, $N_2$, and $N_1$ 
for D4, D3, D2, and D1 rules, respectively.  
The examples of the replacement experiments are shown in
Fig.\ref{replace1}.\\
\renewcommand{\arraystretch}{0.9}
\begin{table}[h]
\begin{center}
\caption{ The numbers of the rules and the replacements for each
figures shown in Fig.\ref{replace1}}
\label{table2} 
\vspace*{0.3cm}
\begin{tabular}{|c|c|c|c|c|c|c|c|c|c|c|c|c|}
     \hline
     \multicolumn{1}{|c|}{Figure} &
     \multicolumn{1}{|c|}{$N_4$}&
     \multicolumn{1}{|c|}{$N_3$}&
     \multicolumn{1}{|c|}{$N_2$}&
     \multicolumn{1}{|c|}{$N_1$}&
     \multicolumn{1}{|c|}{RP4}&
     \multicolumn{1}{|c|}{RP3}&
     \multicolumn{1}{|c|}{RP2}&
     \multicolumn{1}{|c|}{RP1}\\
     \hline
        Fig.\ref{replace1}(a)  &3 &22 &53 &96 &0 &0 &0 &0\\
     \hline
        Fig.\ref{replace1}(b)  &0 &14 &53 &96 &3 &8 &0 &0\\
     \hline
        Fig.\ref{replace1}(c)  &0 &13 &53 &96 &3 &9 &0 &0\\
     \hline
        Fig.\ref{replace1}(d)  &0 &12 &53 &96 &3 &10 &0 &0\\
     \hline
        Fig.\ref{replace1}(e)  &1 &7 &53 &96 &2 &15 &0 &0\\
     \hline
        Fig.\ref{replace1}(f)  &1 &6 &53 &96 &2 &16 &0 &0\\
     \hline
        Fig.\ref{replace1}(g)  &2 &1 &53 &96 &1 &21 &0 &0\\
     \hline
        Fig.\ref{replace1}(h)  &2 &0 &53 &96 &1 &22 &0 &0\\
     \hline
\end{tabular}  
\end{center}  
\vspace{0.5cm}
\vspace{0.5cm}
\end{table}
\indent
The rule table of Fig.\ref{replace1}(a) is obtained randomly with probability
$\lambda=0.6$. At $\lambda=0.6$ most of the randomly created rule tables
generate chaotic patterns. Fig.\ref{replace1}(b) is obtained by 
the 3 RP4s and 8
RP3s, and the numbers of the rules become $N_4=0$ and $N_3=14$,
respectively. These numbers are summarized in Table \ref{table2}.
By this replacements, Fig.\ref{replace1}(b) shows an edge of chaos behavior.
When $N_4=0$, until $N_3=22$ to $N_3=15$, the rule tables generate
chaotic patterns.
Fig.\ref{replace1}(c) is obtained by one more RP3 replacements for
 Fig.\ref{replace1}(b) rule table. It shows a
periodic pattern with a rather long transient length. One more
replacement of RP3 for the Fig.\ref{replace1}(c) rule table is shown in
Fig.\ref{replace1}(d),  where the transient length
becomes shorter.\\
\indent 
Similar replacement experiments for $N_4=1$ and
$N_4=2$ cases are shown in Fig.\ref{replace1}(e), Fig.\ref{replace1}(f), and
Fig.\ref{replace1}(g), Fig.\ref{replace1}(h), respectively.
In these cases, until $N_3 \ge 8$ and $N_3 \ge 2$, rule tables generate
class III patterns at $N_4=1$ and $N_2=2$, respectively. 
In these examples, edges of chaos are observed between classes III and
II behaviors.\\  
\indent  
For each $\lambda=0.8$, $0.75$, $0.7$, $0.6$, $0.5$, $0.4$, $0.3$ 
and $0.2$ point, we have carried out 
a several hundreds to a few thousands replacement experiments.
In these replacements, we have
succeeded in changing chaotic rule tables to a periodic ones by the
replacements of RP4 to RP2, and vice versa by RC4 to RC2.
For the replacements RP4, RP3 and RP2, we have observed no example that 
the rule
table moves to chaotic direction. The converse is true for RC4, RC3 and RC2.
Exceptions are observed only in the replacements RP1 and RC1, which
will be discussed in section 4. \\

\subsection{Chaotic and periodic limit at fixed $\lambda$ in
N-neighbor and K-state CA}
Let us study the effects of the replacements theoretically in the
N-neighbor and K-state CA.
In this general case too, the rule tables are classified into
four groups.
We denote the number of the group 1 rules as $N(g1)$, and similarly for the
numbers of the other groups. 
These numbers satisfy the following sum rules.
\begin{equation}
\begin{array}{ll}
N(g2)+N(g3)=K^{N}(1-\lambda),  \hspace {0.5cm}
N(g1)+N(g4)=K^{N}\lambda, \\
N(g1)+N(g2)=K^{N-1}, \hspace {1.4cm}
N(g3)+N(g4)=K^{N-1}(K-1).
\label{sum_rule1}
\end{array}
\end{equation}
\indent
The individual numbers $N(gi)$ are determined by the probability 
$\lambda$. They are summarized in the Table \ref{N-K}.
They suffer from fluctuations due to 
random number, however in this subsection, we neglect the fluctuations.\\
\renewcommand{\arraystretch}{0.9}
\begin{table}[h]
\caption{ The classification of N-neighbor and K-state CA rules into
four groups. The $\mu_{i}$ represent 0 to K-1 and $h$,$i$ and $l$,
1 to K-1.}
\vspace*{0.3cm}
\label{N-K}
\begin{center}
\begin{tabular}{|c|c|c|c|c|c|c|}
     \hline
     \multicolumn{1}{|c|}{ } &
     \multicolumn{1}{|c|}{rule } &
     \multicolumn{1}{|c|}{N(gi)}\\
     \hline
        group 1  &$T(\mu_1,\mu_2,...,0,...,\mu_{N})=h$  &$K^{N-1}\lambda$\\
     \hline
        group 2  &$T(\mu_1,\mu_2,...,0,...,\mu_{N})=0$  &$K^{N-1}(1-\lambda)$\\
     \hline
        group 3  &$T(\mu_1,\mu_2,...,i,...,\mu_{N})=0$  &$K^{N-1}(K-1)(1-\lambda)$\\
     \hline
        group 4  &$T(\mu_1,\mu_2,...,i,...,\mu_{N})=l$  &$K^{N-1}(K-1)\lambda$\\
     \hline
\end{tabular}  
\end{center}  
\end{table}
\vspace{0.5cm}
\indent
The replacements to decrease the number of the group 1 rule 
while keeping the $\lambda$ fixed are given by,
\begin{equation}
\begin{array}{ll}
N(g1) \rightarrow N(g1)-1,\hspace {0.5cm} N(g2) \rightarrow N(g2)+1\\
N(g3) \rightarrow N(g3)-1,\hspace {0.5cm} N(g4) \rightarrow N(g4)+1\\
\label{del_num_RP}
\end{array} 
\end{equation}
\indent
In the case of 5-neighbor and 4-state CA, they correspond to 
RP4 to RP1.\\
\indent
These replacements stop either when $N(g1)=0$ or $N(g3)=0$ is reached.
Therefore when $N(g1) \le N(g3)$, which corresponds to $\lambda \le
(1-\frac{1}{K})$ in $\lambda$,
all the group 1 rules are replaced by the group 2 rules. 
In this limit, quiescent
states at time $t$ will never be changed, because there is no rule
which 
converts them to other states, while the group 3 rules have a chance to
create a new quiescent state in the next time step. Thus
the number of quiescent states at time t is a non-decreasing function
of t;
therefore, the pattern class should be class I (homogeneous) or 
class II (periodic), which we call periodic limit. Therefore
the replacements of Eq.\ref{del_num_RP} push the rule table toward the
periodic limit.\\ 
\indent
 Let us discuss the reverse replacements of Eq.\ref{del_num_RP}.
In these replacements, if $N(g2) \le N(g4)$, 
all group 2 rules are replaced by
the group 1 rules, except for the quiescent condition. 
In this extreme reverse case, 
all the quiescent states at time t are
converted to other states in next time step, while group 3 rules
create them at different places. 
Then this will most probably develop into chaotic patterns. This limit
will be called chaotic limit, which is reached in the region 
$\frac{1}{K} \le \lambda$.
We should like to say that 
atypical rule table and initial condition might
generate a periodic patterns even in this limit. But in this article, these
exceptional cases are neglected.\\
\indent
Therefore in the following region,
\begin{equation} 
\frac{1}{K} \le \lambda \le 1-\frac{1}{K},
\label{2-limit}
\end{equation}
all the rule tables are located between
these two limit, and by the
replacements of Eq. \ref{del_num_RP} and their reverse ones, the changes 
of the pattern classes take place without fail.
This explains the validity of the hypothesis
of previous section. 
And we have found a method to control the pattern classes at
fixed $\lambda$.

\section{Phase diagram of 5-neighbor and 4-state CA in $\lambda$-F plane}
\begin{figure}
\begin{center}
\scalebox{0.9}{ { \includegraphics{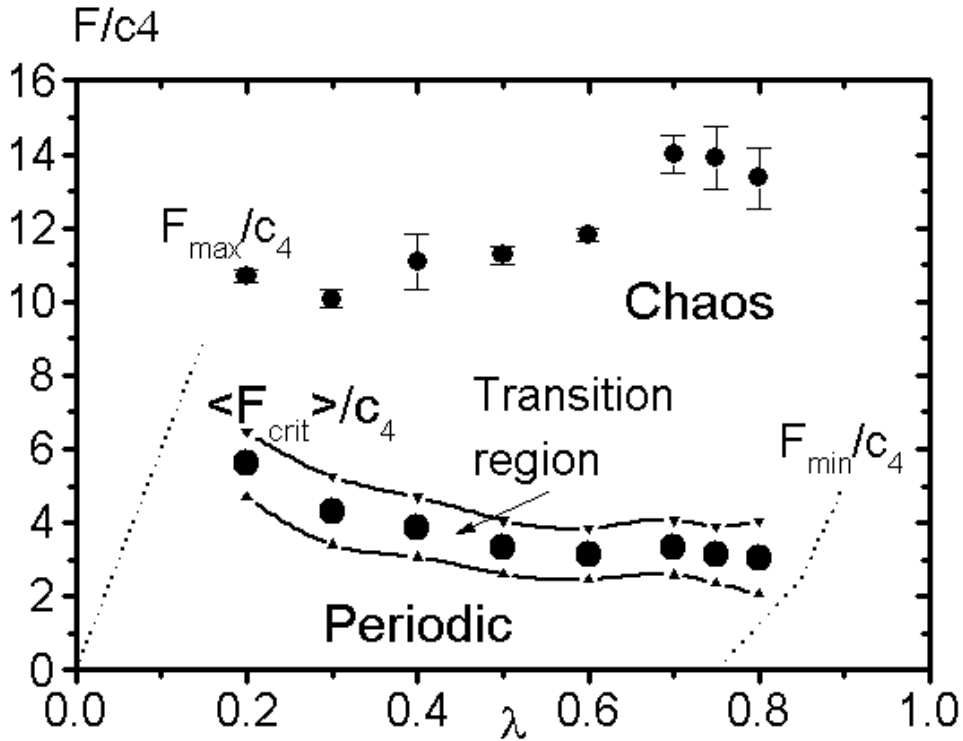} } }
\caption{
Phase diagram of the 5-neighbor and 4-state cellular automata. 
The $<F_{Crit}>/c_{4}$ and the $F_{max}/c_{4}$ are shown with filled circle.
The
$<F_{crit}>/c_{4} + \sigma$ is shown with down triangle and the 
$<F_{crit}>/c_{4} - \sigma$, with up triangle. They are joined by lines.
A schematic sketch of $F_{min}$ in $0.75 \le \lambda$, and $F_{max}$ in
$\lambda \le 0.25$ are also shown by the 
dotted lines. They are drawn based on the arguments in the text.
}
\label{phase_diagram}
\end{center}
\end{figure}

In the previous section, we have found that rule table is located
somewhere between chaotic limit and periodic limit,
in the region $\frac{1}{K} \le \lambda \le 1-\frac{1}{K}$. 
In order to express
the position of the rule table quantitatively, we introduce new
parameter F, which provides us with a new axis ($F$-axis) orthogonal to
$\lambda$.  Minimum of $F$ is the periodic limit, while maximum of it
corresponds to chaotic limit. In this section, we
determine the parameter $F$, for 5-neighbor and 4-state CA.\\
\indent
As a first approximation, the parameter $F$ is taken to be 
be a function of
the numbers of the rules D4, D3, D2 and D1, which have been denoted as
$N_4$, $N_3$, $N_2$ and $N_1$, respectively.  
We proceed to
determine $F(N_4,N_3,N_2,N_1)$ 
by applying simplest approximations and assumptions\\
\indent
We have observed in the replacement experiments, that the position of the
rule table in $F$ moves toward chaotic direction, when $N_4$ or $N_3$ or $N_2$
increases. Examples are shown in Fig. \ref{replace1} and Table
\ref{table2}. However for D1, replacements RP1 and RC1 sometimes look
like random walk on $F$-axis, around the region where the transition of
the pattern class is taken place\footnote{
The effect of D1 rule is to change an isolates quiescent state to
other states in the next time step. This effects 
may easily be compensated by the creation of 
quiescent states by group 3 rules in one time step. This may 
be a reason that 
the replacements of RP1 and RC1 some times look like random walk.}.
Therefore $F$ will be a complicated function of $N_1$ and determination
of it will be difficult.\\ 
\indent
 However the number $N_1$ is rather large, 
therefore we apply mean field approximation for $N_1$.
We replace $N_1$ by its average, and measure $F$ from this
background; namely the $N_1$ dependence is dropped from $F$ and take 
$F=0$ at $N_4=N_3=N_2=0$.\\
\indent
 Here, we apply Taylor series expansion for $F$ at this point, and
approximate it by the linear terms in $N_4$, $N_3$ and $N_2$. \\
\begin{equation}
 F(N_4,N_3,N_2) \simeq c_4 N_4 + c_3 N_3 + c_2 N_2 
 \label{taylor}.
\end{equation} 
where $c_4= \partial F/\partial N_4$, similar for $c_3$ and $c_2$.
They represent the strength of the effects of the rules
D4, D3, and D2 to push the rule table toward chaotic direction.
This definition is symbolic, because $N_4$ is discrete.\\
\indent
 The measure in the $F$ is still arbitrary. We fix it in the unit
where the increase in one unit of $N_4$ results in the  change of $F$ in one
unit. This corresponds to divide $F$ in Eq.\ref{taylor} by $c_4$, and to
express it by the ratio $c_3/c_4$ ($r_3$) and $c_2/c_4$ ($r_2$).\\
\indent
Before we proceed to determine $r_3$ and $r_2$, let us interpret the
parameter $F$ geometrically. Most generally, the rule tables are
classified in 1024-dimensional space in this CA. The rule tables, at
the boundary of the class III and class II
pattern classes at fixed $\lambda$, form 
a hyper-surface in this space. 
We map the points on hyper-surface 
into 3-dimensional $(N_4,N_3,N_2)$ space.
They will be located in some region in the
3-dimensional space. We introduce a surface $F(N_4,N_3,N_2)=\Phi$
in order to line up these points. $F$-axis is a normal line
of the surface $F(N_4,N_3,N_2)=\Phi$.
 In Eq.\ref{taylor}, we
approximate it by a plane. \\
\indent
Our strategy to determine $r_3$ and $r_2$ 
is to find the regression plane in $N_4$, $N_3$, and $N_2$ space. 
It is equivalent to fix the $F$-axis in such a way that
the projection of the distribution of transition points on $F$-axis,
($F_{crit}$) looks as narrow as possible. 
The quality of our approximations and
assumptions reflects the width of the distribution of 
$F_{crit}$.\\
\indent
In the least square method, our problem is formulated to find $r_3$
and $r_2$, which minimize the quantity,
\begin{equation}
s(r_3,r_2)= 
\frac{1}{c_{4}^{2}}\sum_{i,j}(F_{crit}^i(N_4^i,N_3^i,N_2^i)
            - F_{crit}^j(N_4^j,N_3^j,N_2^j))^{2} \label{ansatz},
\end{equation}
where $i$ and $j$ label the rule tables on the hyper-surface 
mapped into 3-dimensional space.
We solve the equations,
$\partial S/\partial r_3=0$ and $\partial S/\partial r_2=0$,
which are
\begin{equation}
\begin{array}{llll}
\displaystyle
 r_3 \sum_{i,j} (\delta N_3^{i,j})^2 +
    r_2 \sum_{i,j} \delta N_2^{i,j}\delta N_3^{i,j}
         =- \sum_{i,j} \delta N_4^{i,j}\delta N_3^{i,j}, \\
\displaystyle
 r_3 \sum_{i,j} \delta N_3^{i,j} \delta N_2^{i,j} +
    r_2 \sum_{i,j} (\delta N_2^{i,j})^2
         =- \sum_{i,j} \delta N_4^{i,j}\delta N_2^{i,j}, \\
\label{sol_r3}
\end{array}
\end{equation}
where $\delta N_4^{i,j}=N_{4}^{i}-N_{4}^{j}$, similar for
$\delta N_3^{i,j}$ and $\delta N_2^{i,j}$.\\
\indent
By artificially carrying out the replacements of RP4, RP3 and RP2 or
RC4, RC3 and RC2,
we look for critical combinations of $N_4$, $N_3$, and $N_2$,
at which the change of the pattern classes are observed.
The examples of the critical combinations are presented in the lines 
of Fig.\ref{replace1}(b),
Fig.\ref{replace1}(e), and
Fig.\ref{replace1}(g) of Table \ref{table2}.\\
\indent
The numbers of the critical combinations ($Nc^{tot}$), which are used
to determine $r_3$ and $r_2$ in Eq. \ref{sol_r3}, are summarized in 
the Table \ref{relative_force}.
The $r_3$ and $r_2$ are determined for each $\lambda$. 
Their results are also shown in Table \ref{relative_force}, 
where the errors are estimated by the jackknife method.\\
\indent
The results show that the coefficients are positive, and satisfy the
order,
\begin{equation}
 c_4 > c_3 > c_2.  
\label{order_force}
\end{equation} 
It means that the effects to move the rule table toward chaotic limit
on the
$F$-axis are stronger for the rules which break longer strings of the
quiescent states. \\
\indent
The order in Eq.\ref{order_force} is understood
by the following intuitive arguments.
If six D4 rules are included in the rule table, the string of the 
quiescent states with length 5 will not develop. Similarly, 
if 33 D3 rules are present in the rule table, no length 
4 string of the quiescent states could be made. These are roughly 
similar situations for pattern classes.
Thus the strength of the D3 rules will be roughly
equal to $6/33$ of that
of D4 rules, similar for the strength of the D2 and D1 rules
\footnote{If we do not impose the quiescent
condition, Eq.\ref{order_force} becomes 
$$c_5 > c_4 > c_3 > c_2.$$  
Therefore the correlation between 
pattern classes and the existence of D5 rule is stronger than the
correlation between those and the number of D4 rules.
If we start our 
study within the quiescent condition, we may make a longer
detour to
find the hypothesis of section 2 and get the qualitative conclusion of
section 3.}.
\renewcommand{\arraystretch}{0.9}\\
\begin{table}[t]
\caption{The number of the critical combinations $Nc^{tot}$ and the 
coefficients $r_3$ and $r_2$. 
\vspace*{0.3cm}}
\label{relative_force}
\begin{center}
\begin{tabular}{|c|c|c|c|c|c|c|}
     \hline
     \multicolumn{1}{|c|}{$\lambda$} &
     \multicolumn{1}{|c|}{  } &
     \multicolumn{1}{|c|}{Relative Strength} &
     \multicolumn{1}{|c|}{Error}&
     \multicolumn{1}{|c|}{$Nc^{tot}$}\\
     \hline
          0.2&    $r_{3}$  &0.1083 &0.0020   &119\\   
&                 $r_{2}$  &0.0153 &0.0025   &  \\
     \hline
          0.3&    $r_{3}$  &0.1083 &0.0028   &90\\ 
&                 $r_{2}$  &0.0069 &0.0021   &  \\
     \hline
          0.4&    $r_{3}$  &0.1182 &0.0074   &58\\ 
&                 $r_{2}$  &0.0165 &0.0072   & \\
 \hline
          0.5&    $r_{3}$  &0.1023 &0.0047   &61\\
&                 $r_{2}$  &0.0262 &0.0010   &  \\
 \hline
          0.6&    $r_{3}$  &0.1288 &0.0019   &101\\
&                 $r_{2}$  &0.0216 &0.0013   &  \\
 \hline
          0.7&    $r_{3}$  &0.1631 &0.0063   &92\\
&                 $r_{2}$  &0.0342 &0.0042   &  \\
 \hline
          0.75&   $r_{3}$  &0.1505 &0.0114   &92\\
&                 $r_{2}$  &0.0419 &0.0065   &  \\
 \hline
          0.8&    $r_{3}$  &0.1315 &0.0168   &89\\
&                 $r_{2}$  &0.0420 &0.0038   &  \\
 \hline
\end{tabular}  
\end{center}  
\end{table}
\begin{figure}
\begin{center}
\scalebox{0.7}{ { \includegraphics{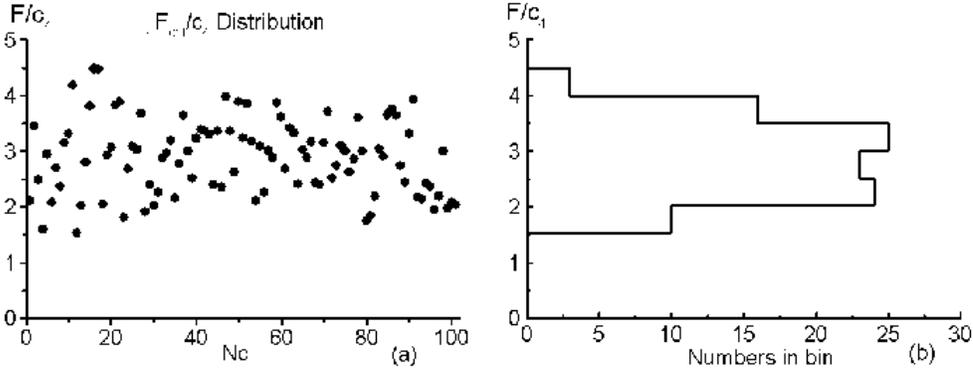} } }
\caption{ Distribution of critical combinations $F_{crit}$ at $\lambda=0.6$.
Fig.\ref{phase_diagram_0.6}a shows 101 individual $F_{crit}$ points.
We label these 101 points by $Nc$.
In Fig.\ref{phase_diagram_0.6}b,
number distribution of the critical combinations in
the bin is displayed. The region of distribution of 
$F_{crit}$ is divided into 6 bins.}
\label{phase_diagram_0.6}
\end{center}
\end{figure}  
\indent
Using these results for $r_3$ and $r_2$, we
calculate $F$s for each critical combination ($F_{crit}$).
For $\lambda=0.6$, they are  
shown in Fig.\ref{phase_diagram_0.6}a for 101 data points.
The number distributions of them are displayed in 
Fig.\ref{phase_diagram_0.6}b. 
From these 101 data, we calculate average, 
$<F_{crit}>$ and standard
deviation $\sigma$ of the distribution of $F_{crit}$.
Similar calculations are carried out for other $\lambda$
points.
The results for $<F_{crit}>$, $<F_{crit}> \pm \sigma$ and maximum of
$F$ ($F_{max}$) determined by the coefficients $r_3$ and $r_2$
are shown in the Fig.\ref{phase_diagram}.
This is a phase diagram in $\lambda-F$ plane.\\
\indent
The phase diagram is natural in the sense that as $\lambda$ increases,
the proportion of the chaotic rule table region increases in the
total $F$ range. 
The two
lines $<F_{crit}> \pm \sigma$ in the figure, indicate that
about $68\%$ of the critical combinations are located in this region, if
the normal distribution is assumed for $F_{crit}$. We call this region
as transition region.  We think that the class II and
class III patterns would not be separated by a line in
$\lambda-F$ plane. Because the transition region in
Fig. \ref{phase_diagram} is a mapping of the hyper-surface, which
separates the two pattern classes at a fixed $\lambda$ in 1024 
dimensional space, into the normal line of the regression plane 
in the 3-dimensional space.
Therefore the two parameters will not be enough to draw the 
boundary by a line.\\
\indent
However we should like to notice that in spite of the simple
approximations and assumptions,
the $F_{crit}$s distribute within a rather narrow portion of total 
$F$ range as shown in Fig. \ref{phase_diagram}.
At $\lambda=0.6$, the total range of 
$F$ is $ 0 \le F \le 11.8$.
All the $F_{crit}$ distribute within the region $1.5 \le F \le
4.48$, which is roughly  $1/4$ of the total region.
The region $<F_{crit}> \pm \sigma$ occupies only about $11\%$ of whole range
of $F$. The similar situation is observed
for the other $\lambda$ points.\\
\indent
The Fig.\ref{phase_diagram} is still qualitative but
rule table space of CA is classified much better by using
$\lambda$ and F. And it
provides us with a deeper understanding of the structure of the CA rule
table space.
The existence of the transition region in $0.2 \le
\lambda \le 0.8$, means that the edge of the chaos could be found 
at least in this region, which is rather wide range in $\lambda$. \\
\indent
Let us discuss the distribution of the rule tables around the
transition region.
All the class IV rule tables are located around transition region 
as Fig. \ref{phase_diagram_0.6} shows,
but converse is not true.
In this region, the three pattern classes coexist and
the number densities of the
class II plus class III rule table are considerably
larger than that of class IV rule tables. 
Outside of the transition region, the probability of finding the class IV
rule table decreases rapidly and 
the pattern classes began to be
classified only by the position in the $\lambda-F$ plane;
in the $<F_{crit}>+\sigma \le F$ region, class III rule tables
dominates
while in the  $F \le <F_{crit}>-\sigma$ region, class II or class I
CA dominates.
The similar distributions of the rule tables are observed at
$\lambda=0.8$, 0.75, 0.7, 0.5, 0.4, 0.3 and 0.2.\\
\indent 
We proceed to the investigation of the end points of the transition region. 
In the subsection 3.2, we have find the region of $\lambda$, where
both chaotic limit and periodic limit coexist.
In the 5-neighbor and K-state CA, it is $1/4 \le \lambda \le 3/4$.
\indent  
In $\lambda < 0.25$ region, not all the group 2 rules could be replaced
by the group 1 rules. Therefore the maximum of $N(g1)$
could not become 256, and it deceases to zero as $\lambda$ approaches to
zero. Then the maximum of $F$, ($F_{max}$) also decreases to
zero toward $\lambda=0$.\\
\indent
Conversely in $\lambda > 0.75$ region, not all the group 1 rules could be
replaced by the group 2 rules. The minimum of 
N(g1) and therefore the minimum of $F$, ($F_{min}$) could not becomes 0.
The line $F_{min}$ increases until $\lambda=1$.
In Fig.\ref{phase_diagram}, we have schematically shown the $F_{max}$ 
and $F_{min}$ lines with dotted lines.\\
\indent
The points where the transition region crosses the $F_{min}$ and
$F_{max}$ lines determine the end points of the coexistence of the two
limits, and also the end points of existence of the class IV
rule tables. 
We have already found that the transition region is found in the region
$0.2 \le \lambda \le 0.8$. Therefore the crossing points are
outside of this region in the case of 5-neighbor 4-state CA. 
\\
\section{Conclusions and discussions}
At a fixed $\lambda$, the patten classes of the CA could 
be controlled by the numbers of the 
group 1 rules, which has been denoted by $N(g1)$.
The maximum of $N(g1)(=K^{N-1}-1)$ corresponds to chaotic limit, 
and $N(g1)=0$
, to the periodic limit.
In the N-neighbor and K-state CA, these two limit exist
in the region $\frac{1}{K} \le \lambda \le 1-\frac{1}{K}$.
Therefore in this
$\lambda$ region, we could control the patten classes by changing $N(g1)$
without fail. The method for it, is the replacements of
Eq. \ref{del_num_RP}.
This provides us with a 
new method to obtain the rule table of edge of chaos.
However, we should like to comment that the fluctuation due to the 
random number will make the statements less rigorous. \\
\indent
This property is studied quantitatively by introducing a new
parameter $F$.
The maximum of the $F$ corresponds to 
the chaotic limit and minimum of it,
the periodic limit. By changing the $F$ using the replacements, we
could find the region, where the transition of
the class II and class III patterns takes place. Thus we could
obtain the phase diagram of the CA in $\lambda-F$ plane.\\ 
\indent
In this article, we have applied the analysis to 5-neighbor and
4-state CA. In this case the group 1 rules are further classified into
5 types as shown in Table \ref{destruc}, and the phase 
diagram is obtained in Fig.\ref{phase_diagram}. 
The transition region 
has a rather gentle
slope as a function of $\lambda$, and it extend at least from
$\lambda=0.2$ until $\lambda=0.8$, which is wider than 
$\frac{1}{4} \le \lambda \le 1-\frac{1}{4}$.  
This explains why the edge of chaos is found in the wide range 
in $\lambda$ for this CA.
It may be interesting whether or not the slope of the transition region 
depends on the models.\\ 
\indent
In the replacement experiments,
we have found the edge of chaos (very long transient lengths) in many
cases. The examples are shown in  Fig.\ref{replace1}.
Sometimes they are
observed in a rather wide range in $N_3$ or $N_2$. 
This indicates that in 
many cases, the transitions are second-order like. But the widths in the
ranges of $N_3$ or $N_2$ are different from each other, and there are
cases where the widths
are less than one unit in the replacement of RP2 (first-order like).
It is very interesting to investigate under what condition
the transition becomes first-order like or second-order like.
The mechanism of the difference in the nature of the transition is an open 
problem and
may be studied by taking into account effects of group 2, 3, and 4
rules. In these studies another new parameters may be found and a more
quantitative phase diagram may be obtained.\\
\indent
It is also interesting to compare the detailed dynamics of class IV CA at
different points of rule table space in $\lambda-F$ plane by applying
Wuencshe's method\cite{wuensche} or by using computational
mechanics\cite{hanson}.\\
\indent
These issues together with finding the  points where the transition
region crosses $F_{max}$ and $F_{min}$ lines in Fig.\ref{phase_diagram},
and the nature of the phase transition at these points will be addressed 
in the forthcoming publications.

\end{document}